\documentclass[aip,apl,reprint]{revtex4-1}
\pdfoutput=1

\usepackage{graphicx}
\usepackage{natbib}
\usepackage{amsmath}

\draft % marks overfull lines with a black rule on the right

\begin{document}




%\preprint{}

\title{Si/SiGe quantum dot with superconducting single-electron transistor charge sensor} %Title of paper








\author{Mingyun Yuan}
\author{Feng Pan}
\author{Zhen Yang}
\author{T. J. Gilheart}
\author{Fei Chen}


%\thanks{}


\affiliation{Department of Physics and Astronomy, Dartmouth College, Hanover, New Hampshire 03755}
\author{D. E. Savage}
\affiliation{University of Wisconsin-Madison, Madison, Wisconsin 53706}
\author{M. G. Lagally}
\affiliation{University of Wisconsin-Madison, Madison, Wisconsin 53706}
\author{M. A. Eriksson}
\affiliation{University of Wisconsin-Madison, Madison, Wisconsin 53706}
\author{A. J. Rimberg}
%\email{Alexander.J.Rimberg@Dartmouth.EDU}
\affiliation{Department of Physics and Astronomy, Dartmouth College, Hanover, New Hampshire 03755}





\date{\today}

\begin{abstract}

We report a robust process for fabrication of surface-gated Si/SiGe quantum dots (QDs) with an integrated superconducting single-electron transistor (S-SET) charge sensor. A combination of a deep mesa etch and AlO$_x$ backfill is used to reduce gate leakage. After the leakage current is suppressed, Coulomb oscillations of the QD and the current-voltage characteristics of the S-SET are observed at a temperature of 0.3 K. Coupling of the S-SET to the QD is confirmed by using the S-SET to perform sensing of the QD charge state. 

\end{abstract}
%\label{}


\maketitle %\maketitle must follow title, authors, abstract and \pacs



\section{Introduction}
Si/SiGe quantum dots (QDs) are promising candidates for quantum computing due to the intrinsically weak spin-orbit interaction in Si, and the existence of the nuclear-spin-free isotope $^{28}$Si. It is therefore expected that $T_1$ and $T_2$ spin relaxation times are longer than those in GaAs.\cite{Tahan:2002p69} Electron-spin resonance has been used to measure these relaxation times. $T_2$ for ensembles of phosphorus donors in Si have been measured to be $\sim14$ ms using conventional microwave measurements, \cite{Tyryshkin:2003p1545} and $\> 100$ $\mu$s using electrical spin trap readout.\cite{Morley:2008} Phenomena such as lifetime-enhanced transport in a Si/SiGe QD have also suggested a long spin relaxation time for individual spins. \cite{Shaji:2008p1969} Recent single shot electrical measurements have found $T_1$ to be $\sim$6 seconds at a field of 1.5 T for phosphorus donors in Si, \cite{Morello:2010p2149} and $\sim$3 seconds at 1.85 T in Si/SiGe QDs. \cite{Simmons:2010p2165} \

There are several reasons for using an S-SET for charge readout in favor of the most common charge sensing scheme, namely a quantum point contact (QPC) in the vicinity of the QD. \cite{Field:1993, Petta:2005p786, Simmons:2009p429, Nordberg:2009} First, the carriers in a QPC are normal electrons and are intrinsically dissipative. In Si/SiGe devices, there is also a typical resistance of a few tens of kiloohms due to the ohmic contacts. An Al SET is superconducting, and shows no dissipation except for that required by the charge sensing process. Furthermore, a QPC is coupled to a dot laterally, whereas the island of the Al SET can be directly on top of the QD. The vertical coupling takes advantage of the large dielectric constants of Si-based materials. Finally, the radio-frequency single-electron transistor (RF-SET),\cite{schoelkopf:1998p640} which has already been used in GaAs based QDs,\cite{Lu:2003p219} has recently been shown to have a combination of high sensitivity (on the order of $10^{-6}e/\sqrt{\text{Hz}}$) and low backaction needed to approach the quantum limit for charge detection. \cite{Brenning:2006p2096,Xue:2007p2078, Xue:2009p2049}\

In order to reliably achieve charge sensing, it is necessary to have a high yield of successful devices. In this paper we introduce a fabrication technique we have developed to produce coupled QD/S-SET systems with higher than 90\% yield, and demonstrate the DC measurement of such a system.\

\section{fabrication}
Fabrication techniques for Si/SiGe quantum devices have developed radically during recent years. Early devices used etching to define dot potentials and side gates.\cite{Klein:2004p1406} Later, Pd Schottky surface gates were adopted to allow more flexible tuning of the QD. Leakage from these gates was suppressed by minimizing the active area of the gate leads and using a deep etch,\cite{Slinker:2005p1984} by fabricating gates with gold sputtering,\cite{Scott:2007p1735} or by growing the Si/SiGe heterostructures using molecular beam epitaxy (MBE).\cite{Berer:2006p1847} Due to the complexity of a coupled SET and gated semiconductor QD device, a high yield of successful devices is critical. Both the Pd dot leads and the Al SET leads need to be leak-free, placing more strict than usual requirements on the surface gates. The fabrication process we have developed to resolve these issues is relatively simple and highly reliable, and could be of use in other applications of Si/SiGe devices requiring extremely high yields.\

The Si/SiGe heterostructure is grown using chemical vapor deposition (CVD). First, a step-graded virtual substrate is grown on Si (001) that was miscut 2 degrees towards (010).  A 1 $\mu$m thick Si$_{0.7}$Ge$_{0.3}$ buffer layer is deposited next, followed by an 18 nm Si well where the two-dimensional electron gas (2DEG) is located. A 22 nm intrinsic layer, a 1 nm doped layer ($\sim10^{-19}$ cm$^{-3}$ phosphorous), a second intrinsic alloy layer of $\sim$ 50 to 76 nm, and last a 9 nm Si cap layer are grown subsequently.\

In order to reduce the leakage current, we use a CF$_4$/O$_2$ plasma in a reactive ion etcher (RIE) to remove the majority of the surface, leaving only the mesa where the QD is formed and the ohmic-contact leads. We then immediately back-fill the etched area with AlO$_x$ in an electron beam (e-beam) evaporator before resist removal, as illustrated by Fig.~\ref{fig1}a.  The etch depth is typically 50 nm beyond the estimated depth of the 2DEG. After an additional patterning step, layered AlO$_x$/Ti/Pd is deposited to form the Schottky gates in the e-beam evaporator. Before gate evaporation we return the sample to the RIE and use CF$_4$ (without O$_2$) to remove the native oxide. Neither the sample surface nor the AlO$_x$ backfill is damaged with this dry etch.\

The QD and the SET are patterned with e-beam lithography. Fig.~\ref{fig1}b shows a scanning electron micrograph (SEM) of a completed QD/S-SET device on a mesa. The dot gates are labeled in the figure. An additional gate helps form a QPC near the QD for back-up charge sensing. The central island of the SET is extended above the QD. After the removal of oxide with CF$_4$, Pd is deposited directly on the mesa to form the dot gates, which are extensions of the photolithographic Pd gates. After QD fabrication is complete, the Al SET and its leads are patterned in a single step and are fabricated with shadow evaporation\cite{Fulton:1987p473} in a thermal evaporator.  Oxygen is introduced and the chamber is kept at 25 mTorr for 2 minutes after the first layer of metal is deposited, creating a thin layer of oxide serving as the tunnel barrier. Before the Al evaporation, any e-beam resist residue is removed by O$_2$ plasma etching. The majority of the Pd gate and SET lead area is located on the oxide. However, the microscopic surface gates used to form the QD and the SET are fabricated directly on the bare Si/SiGe heterostructure.\

\section{Measurement and results}
Samples are cooled to a base temperature of 0.3 K in an Oxford Heliox $^3$He refrigerator. Copper/stainless steel powder filters in the cryostat and $\pi$-type filters at room temperature are used to reduce high-frequency noise. The device under measurement is biased by a dc voltage, sometimes with a small additional ac signal. The conductance of the QD is measured with standard lock-in techniques, and the dc \textit{I-V} characteristics are measured for the SET. Homemade low-noise current and voltage amplifiers are used to amplify the signal. \

To detect the leakage, voltage is applied on each gate and any resulting current through an ohmic contact is measured. Our gate fabrication techniques significantly suppress leakage currents. The Pd gates show no signs of leakage within the sensitivity of our measurement ($\sim$pA) up to an applied voltage of $-3$ to $-5$ V (Fig.~\ref{fig2}a). Without the oxide, leakage currents can become significant before the QPCs can pinch off, preventing the formation of a stable QD in some cases. The back filling of the mesa etch is critical not only for the Pd dot gates but also for the Al SET. In some samples the surface of the oxide is below the mesa (Fig.~\ref{fig2}b). Subsequently the Al leads to the SET are in contact with the mesa edge. In this case, the SET shows no signs of a high-impedance subgap region (Fig.~\ref{fig2}c). We conclude that the high gap currents are a result of the leakage current at the interface of Al and the edge of the mesa (Fig.~\ref{fig2}b). Apparently, the tolerance for leakage of an SET is significantly smaller than that of Pd Schottky gates. To circumvent this problem, we completely seal the edge of the mesa with oxide (Fig.~\ref{fig2}d). In samples fabricated following this procedure, the leakage is further reduced and the superconducting gap of $\sim$1.5 mV is clearly visible in the S-SET \textit{I-V} characteristics(Fig.~\ref{fig2}e).\

\begin{figure}

\includegraphics{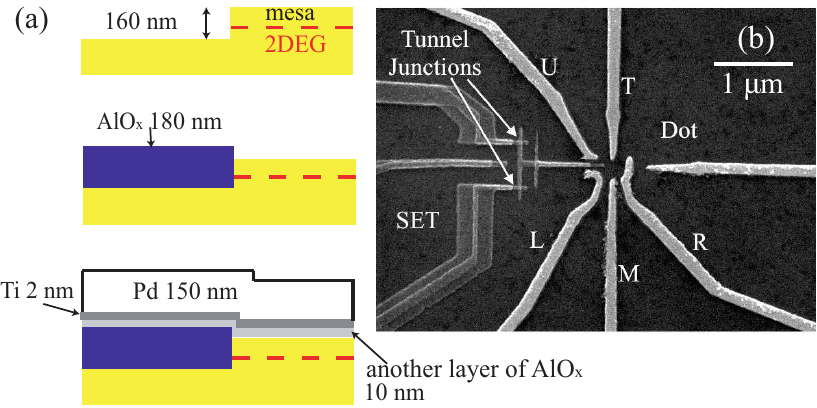}

 \caption{\label{fig1} (Color online) (a) Sequence of fabrication steps used to reduce leakage. From top to bottom: etching, oxide deposition, Pd leads deposition. (b) Scanning electron micrograph of a QD/S-SET device, showing the QD Schottky gates and S-SET fabricated by electron beam lithography. }

 \end{figure}

\begin{figure}

\includegraphics{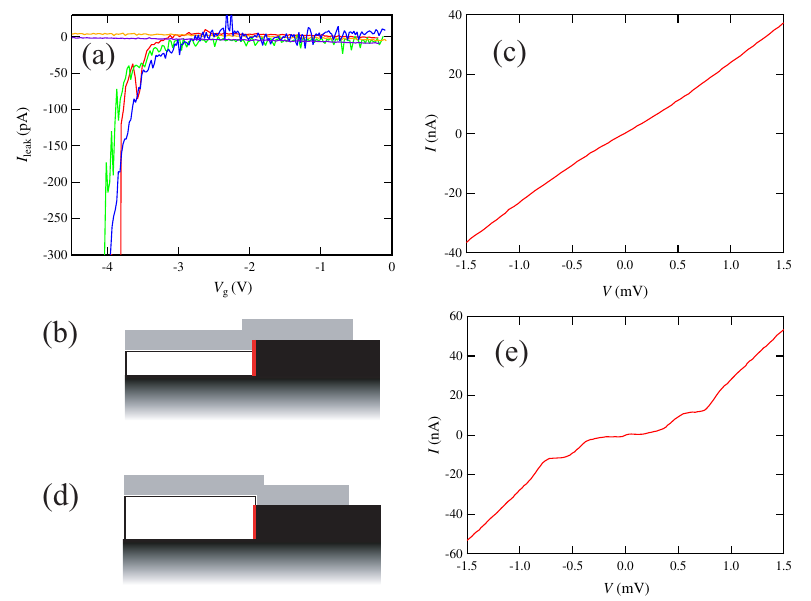}

 \caption{\label{fig2} (a) Leakage current vs. gate voltage for a device fabricated using low-leakage Schottky gates. All the gates remain leak-free up to $-3$ V. (b) Schematic of the mesa edge, showing the mesa (dark), the oxide (white), and Al leads (grey). The mesa edge is highlighted with red. (c) \textit{I-V} curve of an SET, for a mesa that was not completely sealed. (d) Schematic of the mesa edge, which is sealed with a thicker oxide layer. (e) \textit{I-V} curve of another SET, for a mesa completely sealed with oxide. }

 \end{figure}

\begin{figure}

\includegraphics{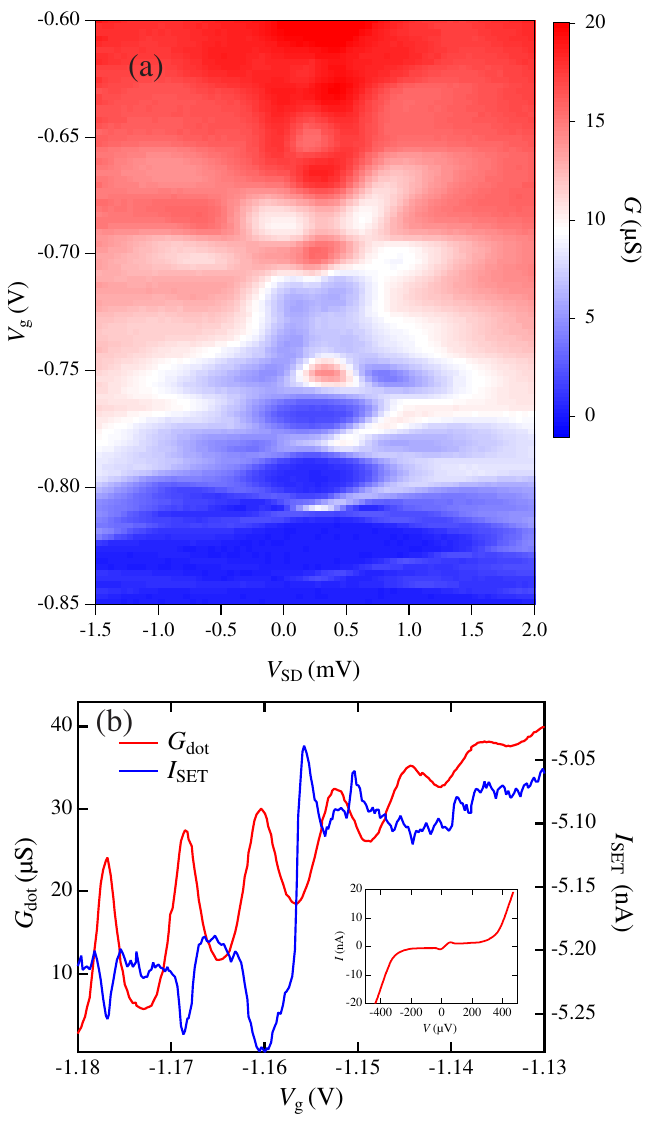}

 \caption{\label{fig3} (a) Differential conductance in a QD vs. bias and gate voltages showing multiple Coulomb diamonds. (b) Simultaneous measurement of SET current and QD conductance demonstrating sensing of the QD charge state.}

 \end{figure}

Once the leakage is eliminated, QDs can readily be formed. Fig.~\ref{fig3}a shows the Coulomb blockade stability plot of the differential conductance of a QD in a sample without an Al SET. The voltages applied on gates R and M (see Fig.~\ref{fig1}b) are $-0.6$ V and $-1.2$ V, respectively. The bias voltage $V_{\text{SD}}$ is swept between $-1.5$ to $2.0$ mV and the voltage $V_{\text{g}}$ of gate T is varied between $-0.85$ to $-0.6$ V. A small DC offset in $V_{\text{SD}}$ is present. Coulomb blockade occurs in the diamond-shaped regions, with possible Kondo effect near $V_{\text{g}}=-0.75$ V, similar to previous reports both in GaAs/AlGaAs \cite{Goldhaber:1998} and Si/SiGe QDs\cite{Klein:2007}. It is estimated from the figure that the gate capacitance $C_{\text{g}}$ is approximately 6 aF and total capacitance $C_\Sigma$ is approximately 46 aF, corresponding to a charging energy of $e^2/C_\Sigma \approx1.7$ meV for the largest diamond (between $V_{\text{g}}=-0.81$ and $-0.84$ V), suggesting that the number of electrons confined in the dot is very small (less than $\sim$10).  It also demonstrates that the Si/SiGe sample has good charge stability.\

Finally, we have also fabricated devices consisting of a QD with an integrated S-SET as in Fig.~\ref{fig1}b. When performing charge sensing, the SET is voltage biased in the subgap region, where it is most sensitive ($V=-0.38$ mV in this case, Fig.~\ref{fig3}b inset). A small change of the island potential will result in rapid variation of the current through the SET. A dot is formed (see Fig.~\ref{fig1}b) with gates T, M, L and U, and gate L is swept. In the voltage range where the dot is well defined, a local minimum of the SET current corresponds to a peak in the QD conductance as in Fig.~\ref{fig3}b. The change of the SET current due to the QD charge state is about 50 pA.\

In conclusion, we have demonstrated that a combination of deep etching and an AlO$_x$ backfill can effectively reduce the leakage current in Si/SiGe heterostructures. An S-SET can be successfully coupled to a Si/SiGe QD, providing the fundamentals for fast real-time charge sensing of an QD with an RF-SET.\

% If in two-column mode, this environment will change to single-column format so that long equations can be displayed. 


%\begin{widetext}

%$$\mbox{put long equation here}$$

%\end{widetext}

% Figures should be put into the text as floats. 

% Use the graphics or graphicx packages (distributed with LaTeX2e).


%



% Put the label that you will use with \ref{} command in the braces of the \label{} command.

%

%

% \begin{table}



% \end{tabular}

% \end{table}


\begin{acknowledgments}

% Put your acknowledgments here.
Fabrication and measurement of samples was supported at Dartmouth and UW-Madison by the NSA, LPS and ARO under agreement no.\ W911-NF-08-1-0482 and at Dartmouth by the NSF under grant number DMR-0804488. Maintenance of the CVD growth facility at UW-Madison has been supported by DOE under Grant No. DE-FG02-03ER46028.  Other facilities support at UW-Madison from NSF/MRSEC, Grant No. DMR-0520527 is acknowledged.  We thank M. Bal, J. Stettenheim and C. B. Simmons for support and discussion.

\end{acknowledgments}


%\bibliography{citation}


\end{document}